\documentclass[conference]{IEEEtran}
\usepackage{cite}\ifCLASSINFOpdf
 
\else
\fi
\usepackage{graphicx}
\usepackage{epstopdf}
\usepackage{epsfig}
\usepackage[cmex10]{amsmath}
\usepackage{amssymb}
\usepackage{amsthm}
\newtheorem{thm}{Theorem}
\newtheorem{propo}{Proposition}
\newtheorem{lem}{Lemma}
\usepackage{array}

\ifCLASSOPTIONcompsoc
  \usepackage[caption=false,font=normalsize,labelfont=sf,textfont=sf]{subfig}
\else
  \usepackage[caption=false,font=footnotesize]{subfig}
\fi
\usepackage{stfloats}
\usepackage{url}
\begin{document}

\title{Computing Quality of Experience of Video Streaming in Network with Long-Range-Dependent Traffic}

\author{\IEEEauthorblockN{Zakaria Ye\IEEEauthorrefmark{1},
Rachid EL-Azouzi\IEEEauthorrefmark{1},
Tania Jimenez\IEEEauthorrefmark{1} and
Yuedong Xu\IEEEauthorrefmark{2}}
\IEEEauthorblockA{\IEEEauthorrefmark{1}University of Avignon, 339 Chemin des Meinajaries, Avignon, France}
\IEEEauthorblockA{\IEEEauthorrefmark{2}Department of Electronic Engineering, Fudan University, China \\
Email: \{zakaria.ye, rachid.elazouzi, tania.jimenez\}@univ-avignon.fr} ydxu@fudan.edu.cn}

\maketitle

\begin{abstract}
We take an analytical approach to study the Quality of user Experience (QoE) for video streaming applications. Our propose  is to characterize buffer starvations for streaming video with Long-Range-Dependent (LRD) input traffic. Specifically we  develop a new analytical framework to investigate Quality of user Experience (QoE) for  streaming   by considering a Markov Modulated Fluid Model (MMFM) that accurately approximates the Long Range Dependence (LRD) nature of network traffic.  We drive the close-form expressions for calculating  the distribution  of starvation as well as start-up delay  using partial  differential equations (PDEs) and solve them using the Laplace Transform.   We illustrate the results with the 
cases of the two-state Markov Modulated Fluid Model that is commonly used in multimedia applications.  We compare our analytical model with simulation  results using  ns-3 under various operating parameters.  We further adopt the
model to analyze the effect of  bitrate switching on the starvation probability and start-up delay. Finally, we  apply  our analysis results to optimize the objective quality of experience (QoE) of media streaming realizing the tradeoff among different metrics incorporating user preferences on buffering ratio, startup delay and perceived quality.

\end{abstract}

\IEEEpeerreviewmaketitle

\section{Introduction}

It has been observed that people addicted to watching streaming videos constitute more than half of the Internet traffic.   With the introduction of smartphones, mobile networks are witnessing  an exponential traffic growth every year. This leads to scenario where Internet and wireless networks are pushed to operate close to their 
performance limits, dictated by current architectural considerations.   Though much effort has been expended and in turn
significant progress has been made in recent years to increase the capacity of mobile networks, there is 
little progress on dealing with the user satisfaction, which is strongly related to the Quality of Experience (QoE).   
This is the big challenge that the operators  face today because they have to look at both the server side and the client side to make a link 
between the quality of service (QoS) of the network and the client satisfaction which depends on the QoE.  Empirical studies in \cite{DOB11, sigcomm13, bala12, krishman12} has identified critical metrics that affect the QoE through the user engagement: 
\begin{itemize}
\item Starvation probability. Denoting the probability that a streaming user sees frozen images.
\item  Average bit-rate. Denoting the mean video quality over the entire session.
\item Bit-rate stability. Describing the jittering of video quality during the entire session.
\item Start-up delay. Denoting the waiting duration between the time that the user requests streaming service and the time that media player starts to play.
\end{itemize}
Paper \cite{DOB11}  pointed out that the buffering ratio is most critical across genres. For example, 1\% increase in buffering reduces 3 minutes for a 90-minutes live video streaming.  \cite{Mok} showed that the total time spent rebuffering and the frequency of rebuffering events have substantial impact on QoE.   Under this context,
media servers and network operators face a crucial challenge on how to avoid the degradation of user perceived media quality based on these metrics.  However, development of  new models as function of these metrics can help operators and content publishers to better invest their network and  server resources toward optimizing these metrics that really matter for QoE.

In this paper we focus on a setting in which a video is streamed over a wireless network which  is subject to a lot of constraints like  bandwidth limitation and rate  fluctuations due  to the frequent changes of channel states and mobility \cite{qoe3}.  Indeed,  time varying network capacity is especially relevant when considering wireless networks where such variations can be caused by fast fading  and slow fading due to shadowing, dynamic interference, and changing loads.  To addresse this issue,  we focus on performance modelling and analysis of a streaming video with Long Range Dependence (LRD) traffic and variable service capacity.   Due to the inherent difficulty and complexity of modelling fractal-like LRD traffic, we assume  that the arrival of packets at player buffer are characterized  by a Markov Modulated Fluid Model, which 
accurately approximates the traffic exhibiting LRD behaviour and mimics the real behaviour of  multimedia traffic with 
short-term and long-term correlation  \cite{mmpp}. In comparison to related works, our whole analysis is on transient regime. 
We construct sets of Partial Differential Equations (PDEs) to derive the starvation probability generating function 
using the external environment, which  is described by the Continuous Time Markov Chain (CTMC).   This approach predicts 
the starvation probability  as function of the file size as well as the prefetching threshold.  Moreover we provide 
relevant results to understand on  how
the starvation probabilities are impacted by the  variation of traffic load and  prefetching
threshold.  We do simulations to show the accuracy of our model using ns-3.   Achieving this goal,  we are able to identify through our model the dependencies between quality metrics. For example,  start-up delay can reduce rebuffing ratio. Similarly bitrate rate switching can reduce buffering. 

With the results developed in this work, we are able to answer the fundamental questions: How many frames  should the media player prefetch to optimize the users' quality of experience?  From what file size the adaptive coding is relevant to avoid the starvation? How bit-rate switching impacts the QoE metrics?   Knowing these answers enables  the user to maximize  his QoE realising the tradeoffs among different metrics incorporating user preferences on rebuffering ratio, start-up delay and quality \cite{NOVA, Beth, Borst}.   We further  introduce an optimization problem  which takes  these key factors in order to achieve the optimal tradeoff between them.  We adopt a more flexible method by defining an objective of QoE  by associated a weight  for each metric based on user preferences \cite{NOVA}.

\section{Related works}
QoE analysis over wireless networks has been studied for many years. In \cite{ballot}, authors study the QoE in a shared 
fast-fading channel using an analytical framework based on Takacs Ballot theorem. They use a GI/D/1 queue to model the 
system, so they assume that the arrival process is independent and identically distributed (i.i.d). In \cite{buffer_starv}, 
the analysis of buffer starvation using M/M/1 queue is performed. They use a recursive method to compare the results with the Ballot 
theorem method even if the recursive method did not offer explicit results. They assume an i.i.d arrival process that is a 
rough model of streaming services over the wireless networks.  Since the performance measures depend on the autocorrelation 
structure of the traffic, a consensus exists  about the limitation of the Poisson process to model the traffic behaviour.   
In \cite{diffusion},  authors  develop an analytical framework to investigate the impact of network dynamics on the user 
perceived video quality, they model the playback buffer by a G/G/1 queue and use the diffusion approximation method to 
compute the QoE. The QoE of streaming from the perspective of the network flow dynamics is studied in \cite{qoe3}. The throughput of a tagged user is governed by the number of the other users in the network.   This study shows 
that the network flow dynamics is the fundamental reason for playback starvation. \\
The rest of this paper is organized as follows: In section \ref{model}, we describe the system model while section \ref{QSM} presents the analysis of the queuing system model. Section \ref{PAQ} describes the performance analysis of the quality of experience and section \ref{2mmpp} presents explicit results for two states MMFM. Section \ref{NA} shows numerical results and section \ref{clc} concludes this paper.

%\noindent***************************\\
%why we choose progressive downloading and not rate adaptation\\
%***************************\\
%Progressive downloading is still an interesting topic, despite the use of the adaptive streaming in many streaming systems, for two reasons. First a lot of streaming web sites did not implement adaptive streaming. Especially the platforms with long duration content, i.e. the movies which duration is around one hour and thirty minutes on average. The second reason is that progressive downloading is implicitly used in adaptive streaming. Indeed, for a given player rate, the system acts like a simple system with the available network bandwidth. \\
%**************************\\

\section{System Model Description}
\label{model}

We consider a single user receiving a media file with finite size $Z$ in streaming. Generally, media files are divided into blocks of frames. When a user makes a request the server segments this media into frames and transfers them to the user through the network (wired and wireless links). 
When frames traverse the internet, their arrivals are not deterministic due to the dynamics of the available bandwidth. 
One of the main characteristics of wireless traffic and Internet traffic in general is the rate fluctuation caused by fast fading and slow fading due to shadowing, dynamic interference, and changing load.  Moreover,  
data packet arrivals in  cellular networks are found to be correlated over both short and long-time
scales. This is generally due to the arrival of packets bursts of comparable size, often leading to high instantaneous arrival rates. Hence, video flows  through the Internet with fluctuating speed.   
In this paper, we assume that frames arrive to the play-out buffer with a rate that can take values from finite set ${\cal S}= \{\lambda_i, i=1,2,..,L\}$.  The rate of arrival frames is governed by a Continuous-Time Markov Chain (CTMC) $ \{ I(t), t \geq 0\} $ with infinitesimal generator $Q$. 
$$
Q =
\begin{pmatrix}
q_{1,1} & q_{1,2} & \cdots & q_{1,L} \\
q_{2,1} & q_{2,2} & \cdots & q_{2,L} \\
\vdots & \vdots & \ddots & \vdots \\
q_{L,1} & q_{L,2} & \cdots & q_{L,L}
\end{pmatrix}
$$
where $ q_{i,i} = -\sum_{j \neq i}{q_{ij}} $.

The maximum buffer size is assumed to be large enough so that the whole file can be stored. %This assumption is justified since all devices have a storage space very large around several GB.  
At the user side, incoming frames are stocked in a buffer and from there they are played with a  rate $\mu$ (e.g., 25 frames per second  (fps)) in the TV and movie-making business.  
We quantify the user perceived media quality using two measures called start-up delay and starvation.
There is ongoing research on mapping these two measures on standard human evaluated QoE measures.
As explained earlier, the media player wants to avoid the starvation by prefetching packets. However, this action might incur a long waiting time. In what follows, we reveal the relationship between the start-up delay and the starvation behaviour, with the consideration of the file size. 

We consider a fluid model that has been proven to be a powerful modeling paradigm in many applications and  relevant to capture the key characteristics that determine the performance  of networks.  Let   
$ r_i = \lambda_i - \mu $ denote  the effective input rate in state $ i $.  Hence the matrix of the effective rates is $ R $, which is a diagonal matrix $ diag \{ \lambda_1 -\mu, \lambda_2 -\mu,...,\lambda_L -\mu \} $.
We denote by $X(t)$ the length of playout buffer of playback at time t. 
 Let $\tau$  be the first  time the buffer is empty before reaching the end of the file, i.e.,   $ \tau = inf\{ t>0: X(t)=0 \} $ and  $T_x$  be the start-up delay where $x$ is the   prefetching threshold.  In the next section we  provide mathematical analysis  to compute the distribution of the number of starvation and start-up delay for a general bursty arrival process.

\section{Analysis of the Queuing System Model}
\label{QSM}
\subsection{Laplace Transform of the Starvation Probability}
\label{first_passage}

We compute the Laplace transform of the probability of starvation given the Continuous-Time Markov Chain $\{I(t), t\geq 0\}$. 
We define $ H_{ij}(x,t) $ to be the probability of starvation in state
$ j $ before time $ t $, given the initial state $ i $ and the initial queue length $ x $.
\begin{equation}
H_{ij}(x,t)=P\{ \tau \leq t, I(\tau)=j | X(0)=x, I(0)=i \}
\end{equation} 
for $ i,j= 1,2,...,L, \quad x>0 $ and $ t \geq 0 $. It is clear that the CTMC cannot be in a state $ j $ at time $ \tau $ if $ r_j > 0 $. Hence $$ H_{ij}(x,t)=0 \quad for \quad all \quad t \geq 0, \quad x \geq 0 \quad if \quad r_j > 0. $$ $$ H_{ij}(x,t)=0 \quad  for \quad all \quad t \geq 0, \quad x > 0 \quad if \quad r_j = 0. $$ $$ H_{ij}(0,t)=1 \quad for \quad all \quad t \geq 0, \quad if \quad r_j = 0. $$  

Let $ \pi = ( \pi_1, \pi_2, ..., \pi_L ) $ be the steady state probability vector of the CTMC $ \{ I(t), t\geq 0 \} $ where $\pi_i$ is the probability to be in the state $i$ at the stationary regime. The expected input and output rates are $ \sum_{i \in S}{\pi_i \lambda_i} $ and $ \sum_{i \in S}{\pi_i \mu_i} $ respectively. The buffer is stable if $ \sum_{i \in S}{\pi_i \lambda_i} < \sum_{i \in S}{\pi_i \mu_i} $. 
Conditioning on the first transition from the state $i$ at time $ 0 $ we have 
\begin{multline}
H_{ij}(x,t) = \sum_{k \neq i}{q_{ik} \Delta t H_{kj}(x+r_i \Delta t, t-\Delta t)} \\
+(q_{ii}\Delta t +1)H_{ij}(x+r_i \Delta t, t-\Delta t)+o(\Delta t)
\end{multline}
Taking the limit $ \lim \limits_{\Delta t \to 0}\frac{H_{ij}(x,t)-H_{ij}(x,t-\Delta t)}{\Delta t} $ and after some algebraic simplification we obtain the following partial differential equation 
%\begin{equation}
%\frac{\partial H_{ij}(x,t)}{\partial t}-r_i\frac{\partial H_{ij}(x,t)}{\partial x}=\sum_{k=1}^{L}{q_{ik}H_{kj}(x,t)}
%\end{equation}
%This can be written in a matrix form as
\begin{equation}
\label{PDE}
\frac{\partial H_{j}(x,t)}{\partial t}-R\frac{\partial H_{j}(x,t)}{\partial x}=QH_{j}(x,t)
\end{equation}
with the initial conditions
\[
H_{ij}(0,t) =
\left \{
\begin{array}{r c l}
1 & if \quad i=j \quad and \quad t \geq 0 \\
0 & if \quad i \neq j \quad and \quad r_i < 0
\end{array}
\right.
\]
\[
H_{ij}(x,0)=0 \quad for \quad all \quad i \neq j \quad and \quad x \geq 0,
\]
\[
H_{jj}(x,0)=0 \quad for \quad x>0.
\]
where $ H_j(x,t)= [H_{1j}(x,t), H_{2j}(x,t),...,H_{Lj}(x,t)] $. \\
The Laplace Stieljes Transform (LST) of $ H_{ij}(x,t) $ is
\begin{eqnarray*}
\tilde{H}_{ij}(x,\omega) &=& \int_{0}^{\infty}{e^{-\omega t}dH_{ij}(x,t)} \\
&=& E[e^{-\omega \tau}; I(\tau)=j | X(0)=x, I(0)=i]
\end{eqnarray*}
for $ i,j = 1,...,L $ and $ \tilde{H}_j(x, \omega) = [\tilde{H}_{.j}(x, \omega)] $.
Taking the LST of Equation (\ref{PDE}) and using the fact that $ H_{ij}(x,0)=0 $ for all $ x > 0 $, we find
\begin{equation}
\label{ODE}
R\frac{d\tilde{H}_j(x,\omega)}{dx}=(\omega I-Q)\tilde{H}_j(x,\omega)
\end{equation}
For a fixed value of $ \omega $, we take $$ \tilde{H}_j(x,\omega) = e^{s(\omega)x}\phi (\omega) $$ as a solution to Equation (\ref{ODE}). Substituting in (\ref{ODE}) we get $$ R s(\omega)\phi (\omega)=(\omega I-Q)\phi (\omega) $$ where the scalar $ s(\omega) $ and the vector $ \phi (\omega) $ are to be determined.
The theorem $ 3.3 $ from \cite{fpt} gives 
\begin{equation}
\tilde{H}_j(x, \omega) = \sum_{s_k(\omega) \in I^-}{ a_{kj} e^{s_k(\omega)x} \phi_{j}^{k}(\omega)}
\label{fp}
\end{equation}
where the coefficients $ a_{kj} $ are obtained by solving 
\[
\sum_{s_k(\omega) \in I^-}{a_{kj} \phi_{j}^{k}(\omega)}=
\left \{
\begin{array}{r c l}
1 & if \quad i=j \\
0 & if \quad i \neq j, r_i \leq 0
\end{array}
\right.
\]
$ s_k(\omega) $ are the roots with negative real parts of $ \Delta(s, \omega) = det(Q+sR-\omega I) $ and $ \phi^k(\omega) $ are the corresponding eigenvectors satisfying the equation 
\begin{equation}
(Q+s(\omega)R-\omega I)\phi(\omega) = 0
\end{equation}

\subsection{Laplace Transform of the Start-up delay}
\label{delayup}

We consider the previous system during the prefetching process and we denote by $X_s(t)$ the length playout buffer of playback at time $t$. Let $$ T_x=inf\{ t \geq 0: X_s(t) \geq x \} $$ be the first time that the length playout buffer  reaches $ x $.
$ T_x $ is the time that the system will take to accumulate $ x $ content in the buffer. This distribution is difficult to solve directly, so we resort to the following duality problem: \\
\textbf{Duality problem:} \textit{What is the starvation probability by time $ t $ if the queue is depleted with rate $ \lambda_i $ and the duration of prefetching contents is $ x $?} \\
This duality problem allows us to compute the prefetching delay as a probability of starvation. We define $ U_{ij}(x,t) $ to be the probability of starvation before time $ t $ at the state $ j $, conditioning on the initial state $ i $ and the initial prefetching content $ x $, i.e., the start-up threshold.
\begin{equation}
U_{ij}(x,t) = P\{ T_x \leq t, I(T_x)=j | I(0)=i, X_s(0)=x \}
\end{equation} 
for $ i,j= 1,2,...,L, \quad x>0 $ and $ t \geq 0 $.

Conditioning on the first transition from the state at time $ 0 $,
\begin{multline}
U_{ij}(x,t) = \sum_{k \neq i}{q_{ik} \Delta t U_{kj}(x-\lambda_i \Delta t, t-\Delta t)} \\
+(q_{ii}\Delta t +1)U_{ij}(x-\lambda_i \Delta t, t-\Delta t)+o(\Delta t)
\end{multline}
Taking the limit $ \lim \limits_{\Delta t \to 0}\frac{U_{ij}(x,t)-U_{ij}(x,t-\Delta t)}{\Delta t} $ and after some algebraic simplification we obtain  the following  partial differential equation 
%\begin{equation}
%\frac{\partial U_{ij}(x,t)}{\partial t}+\lambda_i \frac{\partial U_{ij}(x,t)}{\partial x}=\sum_{k=1}^{L}{q_{ik}U_{kj}(x,t)}
%\end{equation}
%This can be written in a matrix form as
\begin{equation}
\label{PDEDelay}
\frac{\partial U_{j}(x,t)}{\partial t}-R\frac{\partial U_{j}(x,t)}{\partial x}=QU_{j}(x,t)
\end{equation}
with the same initial conditions as in section \ref{first_passage}, 
where $ R=diag\{-\lambda_1, -\lambda_2,...,-\lambda_L \} $ and $ U_j(x,t)= [U_{1j}(x,t), U_{2j}(x,t),...,U_{Lj}(x,t)] $. \\
\begin{equation}
\tilde{U}_j(x, \omega) = \sum_{s_k(\omega)}{ a_{kj} e^{s_k(\omega)x} \phi_{j}^{k}(\omega)}
\label{fpdelay}
\end{equation}
where the coefficients $ a_{kj} $ are obtained by solving 
\[
\sum_{s_k(\omega)}{a_{kj} \phi_{j}^{k}(\omega)}=
\left \{
\begin{array}{r c l}
1 & if \quad i=j \\
0 & if \quad i \neq j
\end{array}
\right.
\]
and $ s_k(\omega) $ are the roots of $ det(Q+sR-\omega I) $ and $ \phi^k(\omega) $ are the corresponding eigenvectors.  

In what follows, we compute  the probability that the prefetching ends at a given state \cite{qoe2}.  For this purpose, we define 
\begin{equation}
V_{ij}(q,x)=P\{ I(T_x)=j | I(0)=i, X_s(0)=q \}
\end{equation}
to be the probability that the prefetching ends at state $j$  given the initial  state i and the initial queue length $q$  where  $x$ is the prefetching  threshold. In the time interval $ [0,h] $, conditioning on the first transition from the state at time $ 0 $, we have 
\begin{multline}
V_{ij}(q,x)=(1+q_{ii}h)V_{ij}(q+\lambda_i h,x) \\
+\sum_{k \neq i}{q_{ik}hV_{kj}(q+\lambda_i h,x)}+o(h)
\end{multline}
After some algebraic simplification and letting $ h \rightarrow 0 $ yields the differential equation
\begin{equation}
\label{ODED2}
diag\{ \frac{1}{\lambda_i}\} \dot{\textbf{V}}(q,x)=-Q\textbf{V}(q,x)
\end{equation}
%\begin{equation}
%\lambda_i \frac{dV_{ij}(q,x)}{dq}=-\sum_{k=0}^{L}{q_{ik}V_{kj}(q,x)}
%\label{ODED}
%\end{equation}
with the boundary condition
\[
V_{ij}(x,x) =
\left \{
\begin{array}{r c l}
1 & \quad if \quad i=j \\
0 & \quad otherwise
\end{array}
\right.
\]
%Equation (\ref{ODED}) can be written in matrix form as
Let $$ Q_v=diag\{ \frac{1}{\lambda_i}\}.(-Q) $$ 
Eq. (\ref{ODED2}) becomes: $ \dot{\textbf{V}}(q,x)=Q_v\textbf{V}(q,x) $.
%\begin{equation}
%\dot{\textbf{V}}(q,x)=Q_v\textbf{V}(q,x)
%\end{equation}
$ \textbf{V}(q,x) $ is given by
\begin{equation}
\label{Sol}
\textbf{V}(q,x)=exp(Q_v q).\textbf{V}(0,x)
\end{equation}
Using Eq.(\ref{Sol}) and the initial conditions, we get
\begin{equation}
\textbf{V}(q,x)=D_v exp(\Lambda_v (q-x))D_{v}^{-1}.\textbf{V}(x,x)
\end{equation}
where $ D_v.\Lambda_v.D_{v}^{-1}=Q_v $, $ \Lambda_v $ is the diagonal matrix containing all the eigenvalues of $ Q_v $ and $ D_v $ is an invertible matrix.

\subsection{The Probability of Starvation and the Start-up Delay}
\label{invert}

In the previous sections we derived explicit expressions for the Laplace-Stieltjes Transform of the probability of starvation and the start-up delay. In this section, we  present  theoretical models to find the corresponding probability of starvation and start-up delay. The Laplace Stieljes Transform of $ H_{ij}(x,t) $ is  
\begin{eqnarray*}
\tilde{H}_{ij}(x,\omega) &=& E[e^{-\omega \tau}; I(\tau)=j | X(0)=x, I(0)=i]\\
&=& \int_{0}^{\infty}{e^{-\omega t}dH_{ij}(x,t)} = \int_{0}^{\infty}{e^{-\omega t}h_{ij}(x,t)}
\end{eqnarray*}
where $h_{ij}$ is the probability density function of $H_{ij}$.
\begin{lem}[Bromwich inversion integral]
Given the Laplace transform $ \tilde{h} $, the function value $ h(t) $ can be recovered from the contour integral 
\begin{equation}
h(t) = \frac{1}{2 \pi i} \int_{b-i \infty}^{b+i \infty}{e^{\omega t}\tilde{h}(\omega) \mathrm d\omega}, \quad t > 0,
\end{equation}
where $ b $ is a real number to the right of all singularities of $ \tilde{h} $, $i^2 = -1$, and the contour integral yields the value $ 0 $ for $ t < 0 $.
\end{lem}
It is shown in \cite{inverse_laplace2} that for real value functions, $ h $ has the following form
\begin{equation}
h(t) = \frac{2 e^{bt}}{\pi}\int_{0}^{\infty}{Re(\tilde{h}(b+iu))cos(ut) \mathrm du}
\end{equation}
According to the \textit{Bromwich inversion integral}, $ h(t) $ can be calculated from the transform $ \tilde{h} $ by performing a numerical integration (quadrature). 
We use a specific algorithm based on the Bromwich inversion integral. It is based on a variant of the Fourier-series method - the trapezoidal rule - 
which proves to be remarkably effective. If we use a step size $ T $, then the trapezoidal rules gives
\begin{multline}
h(t) \approx h_T(t) \equiv \frac{Te^{bt}}{\pi} Re(\tilde{h}(b)) \\
+ \frac{2Te^{bt}}{\pi} \sum_{k=1}^{\infty}{Re(\tilde{h}(b+ikh))cos(kht)}
\end{multline}
where $ Re(\tilde{h}(b))= \tilde{h}(b) $ since $ b $ is real. 
Replacing $ \tilde{h}(\omega) $ by $ \frac{\tilde{H}_{ij}(x, \omega)}{\omega} $ which is the Laplace transform of $ H_{ij}(x,t) $,  
we get the probability of starvation before time $ t $
\begin{multline}
\label{fourier}
P\{ \tau \leq t \}= \frac{2Te^{bt}}{\pi}\bigg[\frac{\tilde{H}_{ij}(x,b)}{2b} \\
+ \sum_{k=1}^{\infty}{Re(\frac{\tilde{H}_{ij}(x,b+ikh)}{b+ ikh})cos(kht)}\bigg]
\end{multline}
The infinite series in ~\eqref{fourier} can simply be calculated by simple truncating because it converges, but more efficient 
algorithm can be obtained by applying a summation acceleration method. An acceleration technique that has proven to be effective 
in our context is Euler summation, after transforming the infinite sum into a nearly alternating series in which successive summands alternate in sign.
We convert \eqref{fourier} into a nearly alternating series by letting $ T=\pi /lt $ and $ b=A/2lt $
%\begin{equation}
%h_T(t) \equiv h_{A,l}(t) = \sum_{k=0}^{\infty}{(-1)^k a_k(t)}
%\end{equation}
\begin{align*}
h_T(t) \equiv h_{A,l}(t) &=\frac{e^{A/2lt}}{2lt}+\frac{2e^{A/2lt}}{lt}\sum_{k=1}^{\infty}{\tilde{h}(\frac{A}{2lt}+\frac{ik\pi}{lt})e^{ik\pi /l}} \\
&= \sum_{k=0}^{\infty}{(-1)^k a_k(t)}
\end{align*}
where
\begin{multline*}
a_k(t)=\frac{e^{A/2l}}{2lt}\Bigg(\tilde{h}(\frac{A}{2lt})1_{\{ k=0\}} \\
+2\sum_{j=1}^{l}{Re\bigg[\tilde{h}(\frac{A}{2lt}+\frac{ij\pi}{lt}+\frac{ik\pi}{t})e^{ij\pi /l}\bigg]}\Bigg)
\end{multline*}
Let $ s_n $ be the approximation $ h_{A,l}(t) $ with the infinite series truncated to $ n $ terms, i.e.,
\begin{equation*}
s_n = \sum_{k=0}^{n}{(-1)^k a_k}
\end{equation*}
where t is suppressed in the notation and $ a_k \equiv a_k(t) $. We apply Euler summation to $ m $ terms after an initial $ n $, so that the Euler sum approximation is 
\begin{equation}
\label{euler}
E(m,n) \equiv E(m,n,t) \equiv \sum_{k=0}^{m}{C_{m}^{k}2^{-m}s_{n+k}}
\end{equation}
Euler summation can be very simply described  as the weighted average of the last $ m $ partial sums by a binomial 
probability with parameter $ m $ and $ p=1/2 $. Hence, \eqref{euler} is the binomial average of the terms $ s_n, s_{n+1}, ..., s_{n+m} $. 
The implementation of the algorithm takes into account the values of $ (l,m,n,A) $. As in \cite{inverse_laplace2}, we use $ (1,M,M,2 ln(10)M/3) $ where $ M=64 $. 
After simplification and letting $ l $ be 1, we get the Euler approximation $ E(t) $ of the inverse $ h(t) $ which seems to be a good approximation
\begin{multline}
\label{eulerfct}
E(t)= \sum_{k=0}^{m}C_{m}^{k}2^{-m}\sum_{q=0}^{n+k}\frac{e^{A/2}}{2 t}\bigg[\tilde{h}(\frac{A}{2t})1_{ \{ q=0 \} } \\
-2Re(\tilde{h}(\frac{A}{2t}+\frac{i\pi (1+q)}{t}))\bigg]
\end{multline}
%\begin{multline}
%\label{pdfH}
%h_{ij}(t)= \sum_{k=0}^{m}C_{m}^{k}2^{-m}\sum_{q=0}^{n+k}\frac{e^{A/2}}{2 t}\bigg[\tilde{H}_{ij}(\frac{A}{2t})1_{\{ q=0\}} \\
%-2Re(\tilde{H}_{ij}(\frac{A}{2t}+\frac{i\pi (1+q)}{t}))\bigg]
%\end{multline}
Eq. (\ref{eulerfct}) looks complicated, but it consists of only $ ((m+1)(m+2n+2))/2 $ 
additions, that is a low computation level. To have the cdf $ H_{ij} $, we just replace $ \tilde{h} $ by $ \frac{\tilde{H}_{ij}(t)}{t} $. The same formula holds for the start-up delay distribution in replacing $ \tilde{h} $ by $ \frac{\tilde{U}_{ij}(t)}{t} $.

\section{Performances Analysis of the Quality of Experience}
\label{PAQ}
In this section we compute the QoE metrics based on the analysis derived  in the previous section. 

\subsection{The Probability of Starvation}

We consider a single user receiving a media file with size $Z$. The necessary time to play 
the whole video if there is no starvation is $ \frac{Z}{\mu } $. Hence, 
using the first passage time distribution $ H_{ij}(x,t) $,  the probability of starvation  happened in state $j$ before reaching the end of file given the initial state  $i$,  is given by 
\begin{multline}
\label{ProSt}
P_s=\sum_{k=0}^{m}C_{m}^{k}2^{-m}\sum_{q=0}^{n+k} e^{A/2}\bigg[\frac{\tilde{H}_{ij}(\frac{\mu A}{2Z})}{ A}1_{\{ q=0\}} \\
-2Re(\frac{\tilde{H}_{ij}(\frac{\mu A+2i\pi \mu (1+q)}{2Z})}{A+2i\pi (1+q)})\bigg]
\end{multline}
The starvation of probability before time $t$ gives an idea of the severity 
of the starvation during the video session.
Let $ D_{ij}(x) := E[\tau , I(\tau)=j | \tau < \infty, I(0)=i, X(0)=x] $ be the mean continuous playback time if the initial state is $ i $, 
the prefetching threshold is $ x $ and the starvation happens in state $ j $.  $ D_{ij}(x) $ is an 
important measure for the severity of starvations. A small $ D_{ij}(x) $ means that the starvation events happen frequently. We find $ D_{ij}(x) $ 
by taking derivatives of $ \tilde{H}_{ij}(x, \omega) $ in $ \omega =0 $.
\[
D_{ij}(x) = -\frac{\partial \tilde{H}_{ij}(x, \omega)}{\partial \omega}|_{\omega =0} \quad i,j=1,...,L
\]  
When the user starts the video session, the initial state is unknown to the system. The video starts 
playing when the prefetching process is finished. Conditioning on the distribution of the entry states $ \pi $, the distribution of the 
states that the playback process begins (or prefetching process ends) is computed by $ \pi .\textbf{V}(0,x) $. Recalling that $ V_{ij}(0,x) $ 
is the probability that the prefetching phase ends at state $ j $ knowing that the video session starts at state $ i $. 
Then the starvation probability with the prefetching threshold $ x $ is obtained by
\begin{equation}
P_s(x)= \pi .\textbf{V}(0,x).\textbf{H}(x,\frac{Z}{\mu })
\end{equation}
where $ \textbf{H} $ is a column vector, $ \textbf{H} =(H_1, H_2,...,H_L)^T $ and $ H_i = \sum_{j=1}^{L}{H_{ij}} $. 
$P_s(x)$ is called the overall starvation probability. The probability of no starvation is $ 1-P_s(x) $.

\subsection{The distribution of the Start-up delay}

The start-up delay is proportional to the start-up threshold. But in the QoE literature, it is more practical to consider the delay rather than the threshold because the delay has a direct impact on the streaming user behaviour. Using the results of the sections \ref{invert} and \ref{delayup}, we derive the cumulative distribution function of the start-up delay
\begin{multline}
U_{ij}(x,t)=\sum_{k=0}^{m}C_{m}^{k}2^{-m}\sum_{q=0}^{n+k} e^{A/2}\bigg[\frac{\tilde{U}_{ij}(\frac{\mu A}{2Z})}{ A}1_{\{ q=0\}} \\
-2Re(\frac{\tilde{U}_{ij}(\frac{\mu A+2i\pi \mu (1+q)}{2Z})}{A+2i\pi (1+q)})\bigg]
\end{multline}
where $ x $ is the start-up threshold, $ Z $ is the file size and $ m, n, A $ are the Euler Summation Algorithm parameters.

\subsection{The generating function of the starvation events}
\begin{figure*}[ht!]
\begin{center}
\includegraphics[scale=0.6]{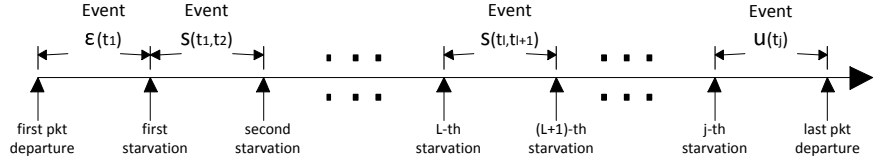}
\caption{\label{StarvL}A path with j starvations}
\end{center}
\end{figure*}
When a starvation event happens, the media player pauses until $ x $ contents are re-buffered. We are interested in the 
probability distribution of the starvations, given the file size $ Z $. We define a \textit{path} as a complete sequence of frames 
arrivals and departures. We illustrate a typical path with $ j $ starvations in Fig. \ref{StarvL}. The path can be decomposed 
into three types of mutually exclusive events as follows:
\begin{itemize}
\item[.] Event $ \mathcal{E} (t_1) $: the buffer becoming empty for the first time in the entire path.
\item[.] Event $ \mathcal{S}_l (t_l,t_{l+1}) $: the empty buffer after the instant $ t_{l+1} $ given that the previous empty buffer happens at $ t_l $.
\item[.] Event $ \mathcal{U}_j (t_j) $: the last empty buffer observed after the instant $ t_j $. 
\end{itemize}
Obviously, a path with $ j $ starvations is composed of a succession of events $$ \mathcal{E}(t_1), 
\mathcal{S}_1(t_1,t_2), \mathcal{S}_2(t_2,t_3),...,\mathcal{S}_{j-1}(t_{j-1},t_{j}),\mathcal{U}_j(t_j) $$ 
We let $ P_{\mathcal{E}(t_1)} $, $ P_{\mathcal{S}_{l}(t_{l},t_{l+1})} $ and $ P_{\mathcal{U}_j(t_j)} $ 
be the probabilities of events $ \mathcal{E}(t_1) $, $ \mathcal{S}_{l}(t_{l},t_{l+1}) $ and $ \mathcal{U}_j(t_j)$ respectively. 
The probability distribution of event $ \mathcal{E}(t_1) $ is expressed as
\begin{equation}
P_{\mathcal{E}(t_1)} =
\left \{
\begin{array}{l l l}
0, \quad \quad if \quad \mu t_1 < x \quad or \quad \mu t_1 \geq Z; \\
\pi.\textbf{V}(0,x).\textbf{h}(x,t_1), \quad otherwise.
\end{array}
\right.
\end{equation}
where $ \textbf{V} $ and $ \textbf{h} $ are $ L $x$ L $ and $ L $x$ 1 $ matrices respectively. 
The first starvation cannot happen at the departure of first $ (x-1) $ contents because of the prefetching of $ x $ contents. 
It cannot happen after all $ Z $ contents have been served because this empty buffer is not a starvation. 
For $ \mu t_1 \in [x,Z [ $ the starvation happens at time $ t_1 $ conditioned on the states that the playback process begins. 
The probability distribution of event $ \mathcal{U}_j(t_j) $ is given by 
\begin{equation}
P_{\mathcal{U}_j(t_j)} =
\left \{
\begin{array}{l l l}
0, \quad \quad if \quad \mu t_j < jx \quad or \quad \mu t_j \geq Z; \\
1, \quad \quad if \quad Z-x \leq \mu t_j < Z; \\
\textbf{V}(0,x).(\textbf{1}-\textbf{H}(x,\frac{Z}{\mu }-t_j)), \quad otherwise.
\end{array}
\right.
\end{equation}
where $ \textbf{H} $ is a column vector. $ t_j $ is the time of the $ j $-th starvation. The extreme case is that these $ j $ 
starvations take place consecutively.Then $ \mu t_j $ should be greater than $ jx $. Otherwise there cannot have $ j $ 
starvations. If $ \mu t_j $ is no less than $ Z-x $, the media player resumes until all the remaining content $ Z-\mu t_j $ 
is stored in the buffer. Then, starvation will not appear afterwards. In the remaining case, it is the probability of having 
no starvation after time $ t_j $. We denote by $ P_s(j) $ the probability of having $ j $ starvations. 
The case with one starvation is given by
\begin{equation}
P_s(1)= \int_{t=0}^{\frac{Z}{\mu }}{P_{\mathcal{E}(t)}.P_{\mathcal{U}_1(t)}} \mathrm dt
\end{equation}
To compute the probability of having more than one starvation, we need to find the probability of event $ \mathcal{S}_{l}(t_{l},t_{l+1}) $. $ \mu t_l $ 
should not be less than $ l x $ in order to have $ l $ starvations. Given that the buffer is empty just after time $ t_l $, the $ (l+1)^{th} $ 
starvation cannot happen at $ \mu t_{l+1} \in [\mu t_l +1, \mu t_l +x-1] $ because of the prefetching process. Since there are $ j $ 
starvations in total, the $ (l+1)^{th} $ starvation must satisfy $ \mu t_{l+1} < Z-(j-l-1)x $. We next compute the remaining case that 
the $ l^{th} $ and the $ (l+1)^{th} $ starvations happen at time $ t_l $ and $ t_{l+1} $ respectively. We compute this probability 
using the first passage time density when the starvation happens at time $ t_{l+1} $ and the initial time was $ t_l $ with a 
prefetching process. $ P_{\mathcal{S}_{l}(t_{l},t_{l+1})} $ is expressed as 
\begin{equation}
\left \{
\begin{array}{c c c}
\textbf{V}(0,x).\textbf{h}(x,t_{l+1}-t_l),  \\ if \quad \mu t_l \geq l x, \mu t_l +x \leq \mu t_{l+1} < Z-(j-l-1)x; \\
0, \quad otherwise.
\end{array}
\right.
\end{equation}
We use in this method a trick that concerns the time scale. Every time the player resumes for the prefetching 
process we resume also the time scale, that means if the starvation happens at time $ t $, the player will start playing 
at the same time $ t $ with $ x $ initial contents in the buffer. The probability of having $ j (j \geq 2) $ starvations is given by
\begin{multline}
P_s(j) = \int_{t_1=0}^{\frac{N}{\mu }}\int_{t_2=0}^{\frac{N}{\mu }} \cdots \int_{t_{j-1}=0}^{\frac{N}{\mu }}
\int_{t_j=0}^{\frac{N}{\mu }}P_{\mathcal{E}(t_1)}.P_{\mathcal{S}_{1}(t_{1},t_2)} \cdots \\
P_{\mathcal{S}_{j-1}(t_{j-1},t_{j})}.P_{\mathcal{U}_j(t_j)} \mathrm dt_1 \mathrm dt_2 \cdots \mathrm dt_{j-1} \mathrm dt_j
\end{multline}
In the next section, we provide explicit expressions of QoE metrics where CTMC has two states. 

\section{The 2-state MMFM Source}
\label{2mmpp}
\begin{figure}[ht!]
\begin{center}
\includegraphics[scale=0.25]{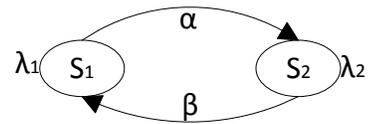}
\caption{\label{2mpfig}The two-state MMPP source}
\end{center}
\end{figure}

In this section, we consider a special case in which the CTMC has two states : $ \{ 1,2 \} $  (see Fig. \ref{2mpfig}) with infinitesimal generator $Q$ and rate matrix $R$ \[
Q=
\begin{pmatrix}
-\beta & \beta \\
\alpha & -\alpha
\end{pmatrix}
\quad \quad
R=
\begin{pmatrix}
\lambda_2 -\mu & 0 \\
0 & \lambda_1 - \mu 
\end{pmatrix}
\]
Our objective is to understand the interaction between the parameters of arrival process and the probability of starvation. Using the results of section \ref{first_passage} irrespective of the condition of stability of the queue $ \pi_1 \lambda_1 + \pi_2 \lambda_2 < \mu $, we get:
\begin{multline}
\Delta(s, \omega) = det(Q+sR-\omega I) \\
=(\lambda_1 -\mu)(\lambda_2 - \mu )s^2 - [(\lambda_1 - \mu )(\omega + \beta) \\ 
+ (\lambda_2 - \mu )(\omega + \alpha)]s + \omega (\omega + \alpha + \beta) 
\end{multline}
It is a polynomial of degree 2 in $ s $ where the two zeros are given by:
\begin{equation}
s_1(\omega) = \frac{b+\sqrt{b^2-4\omega (\omega + \alpha + \beta)(\lambda_1 - \mu )(\lambda_2 - \mu )}}{2(\lambda_1 - \mu)(\lambda_2 - \mu )}
\end{equation}
\begin{equation}
s_2(\omega) = \frac{b-\sqrt{b^2-4\omega (\omega + \alpha + \beta)(\lambda_1 - \mu )(\lambda_2 - \mu )}}{2(\lambda_1 - \mu)(\lambda_2 - \mu )}
\end{equation}
where $ b=(\lambda_1 - \mu )(\omega + \beta) + (\lambda_2 - \mu )(\omega + \alpha) $. Equation~\eqref{fp} contains terms with only $ Re(s_k(\omega)) < 0 $. So we have to determine the signs of $ Re(s_1(\omega)) $ and $ Re(s_2(\omega)) $. The next propositions give the placement of these two zeros in the complex plane.
\begin{propo}
Let $ \lambda_1 > \lambda_2 $,
\begin{itemize}
\item[1.] $ \lambda_2 > \mu $ $ \Rightarrow $ $ \lambda_1 > \mu $ , so $ \lambda_1 - \mu > 0 $ and $ \lambda_2 - \mu > 0 $ then no starvation.
\item[2.] $ \lambda_2 < \mu $ and $ \lambda_1 > \mu $ , so $ \lambda_2 - \mu < 0 $ and $ \lambda_1 - \mu > 0 $ then $ Re(s_2(\omega)) > 0 $ and $ Re(s_1(\omega)) < 0 $.
\item[3.] $ \lambda_2 < \mu $ and $ \lambda_1 < \mu $ , so $ \lambda_2 - \mu < 0 $ and $ \lambda_1 - \mu < 0 $ then $ Re(s_2(\omega)) < 0 $ and $ Re(s_1(\omega)) < 0 $.
\item[4.] $ \lambda_2 = \mu $ $ \Rightarrow $ $ \lambda_1 > \mu $ then no starvation.
\item[5.] $ \lambda_1 = \mu $ $ \Rightarrow $ $ \lambda_2 < \mu $, $ s_2(\omega) = s_1(\omega) =s(\omega)= \frac{\omega (\omega + \alpha + \beta)}{(\lambda_2 - \mu )(\omega + \alpha)}$ and $ Re(s(\omega)) < 0 $.
\end{itemize}
\end{propo}
\begin{propo}
Let $ \lambda_2 > \lambda_1 $,
\begin{itemize}
\item[1.] $ \lambda_1 > \mu $ $ \Rightarrow $ $ \lambda_2 > \mu $ , so $ \lambda_1 - \mu > 0 $ and $ \lambda_2 - \mu > 0 $ then no starvation.
\item[2.] $ \lambda_1 < \mu $ and $ \lambda_2 > \mu $ , so $ \lambda_1 - \mu < 0 $ and $ \lambda_2 - \mu > 0 $ then $ Re(s_2(\omega)) > 0 $ and $ Re(s_1(\omega)) < 0 $.
\item[3.] $ \lambda_1 < \mu $ and $ \lambda_2 < \mu $ , so $ \lambda_1 - \mu < 0 $ and $ \lambda_2 - \mu < 0 $ then $ Re(s_2(\omega)) < 0 $ and $ Re(s_1(\omega)) < 0 $.
\item[4.] $ \lambda_1 = \mu $ $ \Rightarrow $ $ \lambda_2 > \mu $ then no starvation.
\item[5.] $ \lambda_2 = \mu $ $ \Rightarrow $ $ \lambda_1 < \mu $, $ s_2(\omega) = s_1(\omega) =s(\omega)= \frac{\omega (\omega + \alpha + \beta)}{(\lambda_1 - \mu )(\omega + \beta)}$ and $ Re(s(\omega)) < 0 $.
\end{itemize}
\end{propo}

\begin{propo}
Let $ \lambda_1 = \lambda_2 = \lambda $,
\begin{itemize}
\item[1.] $ \lambda > \mu $, no starvation.
\item[2.] $ \lambda < \mu $, $ Re(s_2(\omega)) < 0 $ and $ Re(s_1(\omega)) < 0 $.
\item[3.] $ \lambda = \mu $, no starvation because of the prefetching.
\end{itemize}
\end{propo}
The LST $ \tilde{H}_j(x,\omega) $ of the distribution is given in the next theorem.
\begin{thm}
\begin{itemize}
\item[1.] When $ \lambda_1 < \mu $, $ \lambda_2 \geq \mu $, $ \tilde{H}_2(x,\omega) =0 $ and
\begin{align*}
\tilde{H}_1(x,\omega) &= \begin{bmatrix}
\tilde{H}_{11}(x,\omega) \\
\tilde{H}_{21}(x,\omega)
\end{bmatrix} 
&= e^{s_1(\omega)x} \begin{bmatrix}
1 \\
\frac{\beta+\omega -(\lambda_2 -\mu )s_1(\omega)}{\beta}
\end{bmatrix}
\end{align*}
\item[2.] When $ \lambda_2 < \mu $, $ \lambda_1 \geq \mu $, $ \tilde{H}_1(x,\omega) =0 $ and
\begin{align*}
\tilde{H}_2(x,\omega) &= \begin{bmatrix}
\tilde{H}_{22}(x,\omega) \\
\tilde{H}_{12}(x,\omega)
\end{bmatrix} 
&= e^{s_2(\omega)x} \begin{bmatrix}
1 \\
\frac{\beta+\omega -(\lambda_2 -\mu )s_2(\omega)}{\beta}
\end{bmatrix}
\end{align*}
\item[3.] When $ \lambda_1 < \mu $, $ \lambda_2 < \mu $,
\begin{align*}
\tilde{H}_1(x,\omega) &= a_{21}e^{s_2(\omega)x} \begin{bmatrix}
1 \\
\frac{\beta+\omega -(\lambda_2 -\mu )s_2(\omega)}{\beta}
\end{bmatrix}+ \\
&~~~~ a_{11}e^{s_1(\omega)x} \begin{bmatrix}
1 \\
\frac{\beta+\omega -(\lambda_2 -\mu )s_1(\omega)}{\beta}
\end{bmatrix}
\end{align*}
\begin{align*}
\tilde{H}_2(x,\omega) &= a_{22}e^{s_2(\omega)x} \begin{bmatrix}
1 \\
\frac{\beta+\omega -(\lambda_2 -\mu )s_2(\omega)}{\beta}
\end{bmatrix}+ \\
&~~~~ a_{12}e^{s_1(\omega)x} \begin{bmatrix}
1 \\
\frac{\beta+\omega -(\lambda_2 -\mu )s_1(\omega)}{\beta}
\end{bmatrix}
\end{align*}
\end{itemize}
where
\begin{align*}
a_{11} &=\frac{\beta+\omega -(\lambda_2 - \mu )s_1(\omega)}{(\lambda_2 -\mu )(s_2(\omega)-s_1(\omega))},
a_{21} =\frac{\beta+\omega -(\lambda_2 -\mu )s_2(\omega)}{(\lambda_2 - \mu )(s_1(\omega)-s_2(\omega))} \\
a_{22} &=\frac{\beta}{(\lambda_2 -\mu )(s_1(\omega)-s_2(\omega))},
a_{12} =\frac{\beta}{(\lambda_2 -\mu )(s_2(\omega)-s_1(\omega))}
\end{align*}
\end{thm}
The proof of this theorem can be find in the appendix of \cite{arxiv}. Taking $ \lambda_2 = 0 $ gives the first passage time distribution for the ON-OFF source.

\section{Numerical Analysis}
\label{NA}
\subsection{Simulation}

\begin{figure}[ht!]
\begin{center}
\includegraphics[scale=0.5, height=5.2cm]{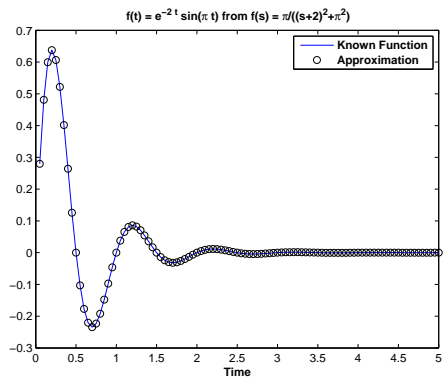}
\caption{\label{testfig}The accuracy of Euler Summation Algorithm}
\end{center}
\end{figure}

\begin{figure}[ht!]
\begin{center}
\includegraphics[scale=0.5, height=5.2cm]{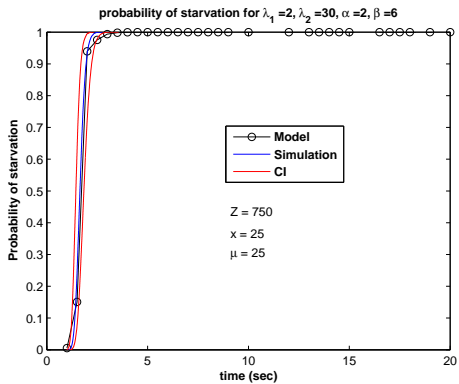}
\caption{\footnotesize{The probability of starvation for two states} \label{OFFinf}}
\end{center}
\end{figure}

We use ns-3 simulator in order to compare the dynamics of the process with our model. The simulation topology consists on a server and a client in order to simulate the queue model. The server sends the traffic to the client following the Continuous Time Markov Chain. The client holds a buffer where the traffic is stored. The parameters of the traffic depend on the CTMC parameters. Then we analyze the behavior of the client buffer content which simulates the player. We run 10 simulations and compute the 95\% confidence interval on all observed metrics, but it is not shown on all the figures for improving readability because it is very narrow.  
We first show the accuracy of the method that we use to invert the Laplace Transform. In Fig. \ref{testfig}, we plot the known inverse Laplace 
Transform of the function $ f(t)=e^{-2t} sin(\pi t) $ that is $ f(s)=\pi /((s+\pi )^2 +\pi^2) $ and the inverse using formula (\ref{eulerfct}) of section \ref{invert}.
Fig. \ref{OFFinf} shows the starvation probability for a two states MMFM source for $ \lambda_1 = 2 $, $ \lambda_2 = 30 $ and $ \mu = 25 $, that means the buffer size increases on state $2$ and decreases on state $1$. 
This is done for states transitions $ \alpha = 2 $ and $ \beta = 6 $. 
%These figures show that our analytical results match the simulations.
\begin{figure}[ht!]
\begin{center}
\includegraphics[scale=0.5, height=5.2cm]{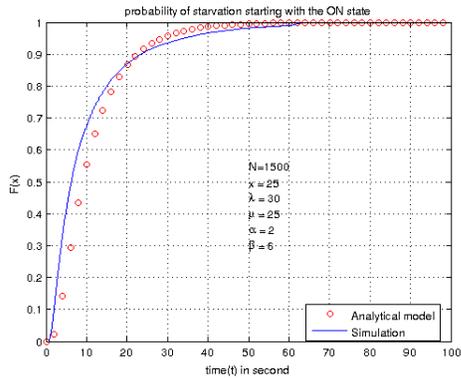}
\caption{\footnotesize{The probability of starvation for $ \lambda > \mu $} \label{ONsup}}
\end{center}
\end{figure}
\begin{figure}[ht!]
\begin{center}
\includegraphics[scale=0.5, height=5.2cm]{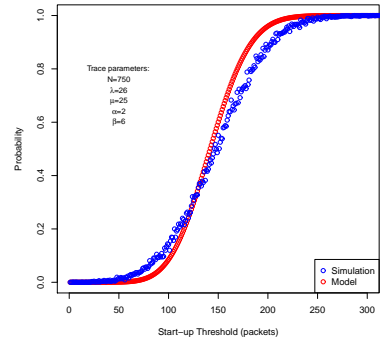}
\caption{\footnotesize{The probability of no starvation versus the start-up threshold x} \label{NoStarvX}}
\end{center}
\end{figure}
\begin{figure}[ht!]
\begin{center}
\includegraphics[scale=0.3]{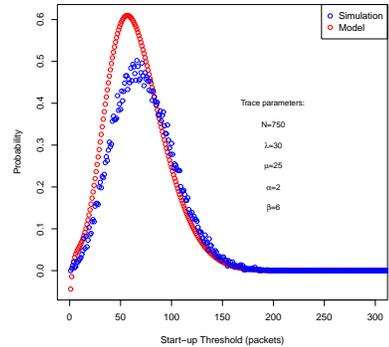}
\caption{\footnotesize{The probability of having one starvation} \label{OneStarv}}
\end{center}
\end{figure}
%\begin{figure}[ht!]
%\begin{center}
%\includegraphics[scale=0.3]{fig8.eps}
%\caption{\footnotesize{The probability of having two starvations} \label{TwoStarv}}
%\end{center}
%\end{figure}
Fig. \ref{NoStarvX} illustrates the impact of the start-up threshold $ x $ on the probability of no starvation.
Fig. \ref{OneStarv} shows the probability of having one starvation.  These   simulation results validate the correctness of our analysis. Hence, in the following experiments, we only illustrate the analytical results.

\subsection{Performances Evaluation}
\begin{figure}[ht!]
\begin{center}
\includegraphics[scale=0.5, height=5.2cm]{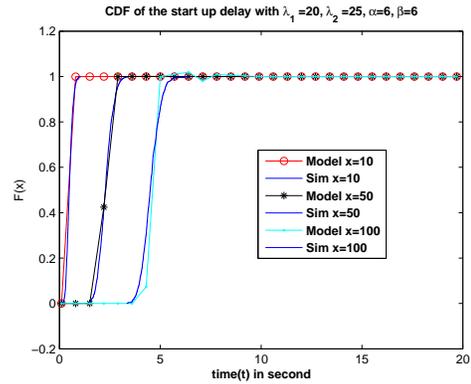}
\caption{\footnotesize{The CDF of the start up delay for the two states mmpp} \label{startupdelay}}
\end{center}
\end{figure}
%\begin{figure}[ht!]
%\begin{center}
%\includegraphics[scale=0.5, height=5.2cm]{no_starvN3.eps}
%\caption{\footnotesize{The probability of no starvation versus the file size N} \label{NoStarvN}}
%\end{center}
%\end{figure}
\begin{figure}[ht!]
\begin{center}
\includegraphics[scale=0.5, height=5.2cm]{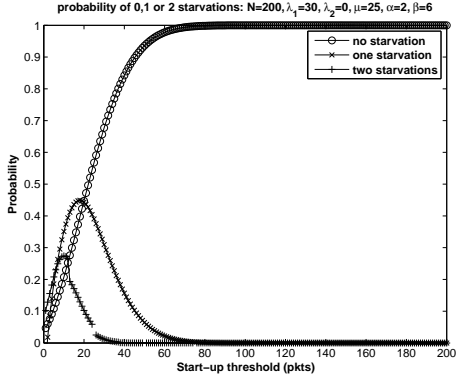}
\caption{\footnotesize{Probability of 0,1 and 2 starvations versus the start-up threshold x} \label{NStarvX}}
\end{center}
\end{figure}
\begin{figure}[ht!]
\begin{center}
\includegraphics[scale=0.5, height=5.2cm]{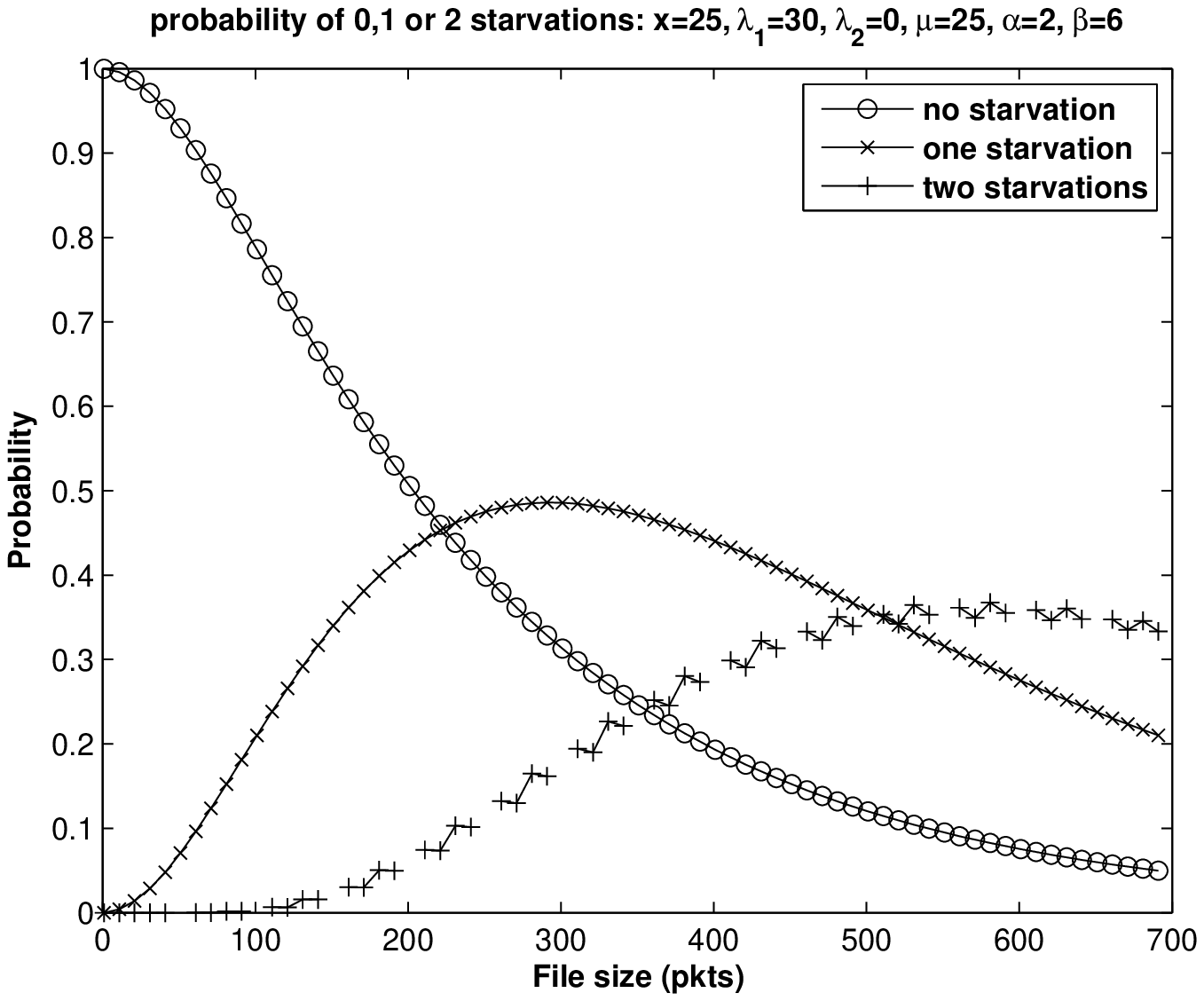}
\caption{\footnotesize{The probability of 0,1 and 2 starvations versus the file size N} \label{NStarvN}}
\end{center}
\end{figure}
Fig. \ref{startupdelay} gives the CDF of the start-up delay for different values of start-up threshold $ x $. We can see that the start-up delay increases with $ x $.  On the other hand figure \ref{NoStarvX} illustrates the impact of the start-up threshold $ x $ on the probability of no starvation. When $ x $ is large enough (near 300 pkts in the figure) no starvation 
will happen until the end of the video. Since, the curve grows sharply, it is clear that a slight increase in $ x $ can greatly 
improve the starvation probability. 
%On the other side, when the file size is large enough (near 400 pkts in the figure), 
%starvation will happen for sure. This curve shows that starvation issue is more severe for long video duration.
In figure \ref{NStarvX}, we plot the probability of having 
no more than two starvations with $ \lambda_1=30 $, $ \lambda_2=0 $, $ \mu =25 $ irrespective to $ x $. When $ x $ is large enough, no starvation will 
happen until the end of the video session. On the other side, figure \ref{NStarvN} shows that the starvation happens for sure when the file size approaches infinity. 
The curves of the probability of having one starvation or two starvations increase first, and then decrease to zero. This means that the starvation can be avoided when $ x $ is large enough. 
The two curves have a maximum value at a given start-up threshold or a given file size. 
So one can choose the threshold $ x $ to have exactly one or two starvations. This is an important measure because it allows one to achieve the buffer requirements in setting up the desired values. Indeed, a very small threshold do not help to reduce the starvation probability and very large thresholds do not further reduce the starvation probability.  Hence the analytical model aims to predict the player buffer behavior in video streaming sessions. The network parameters and the video size are the framework inputs that can be used to improve the QoE related to user preferences.

\subsection{Optimization of the QoE}
\begin{figure}[ht!]
\begin{center}
\includegraphics[scale=0.5, height=5.2cm]{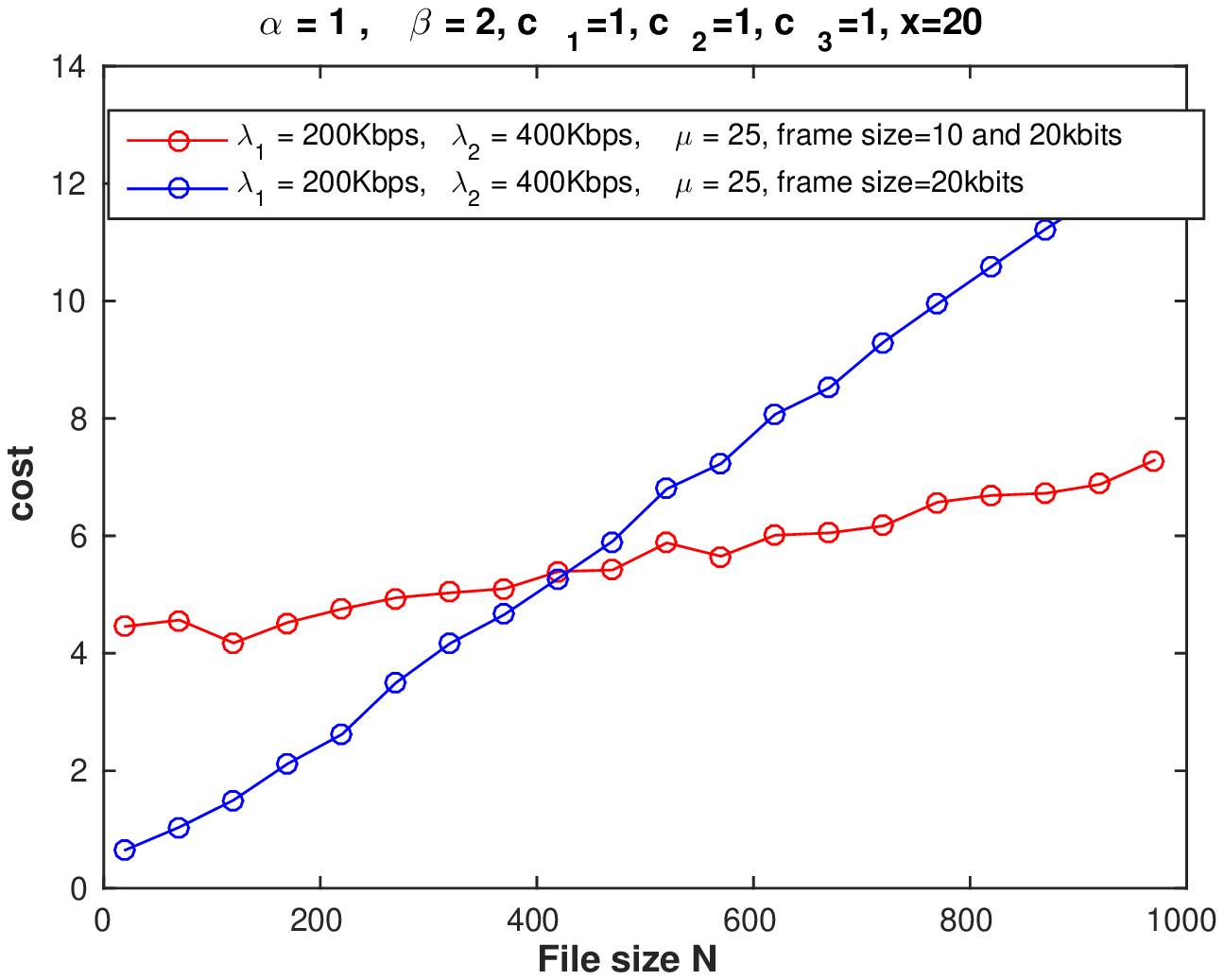}
\caption{\footnotesize{The cost for progressive streaming versus the adaptive streaming} \label{progadap}}
\end{center}
\end{figure}

\begin{figure}[ht!]
\begin{center}
\includegraphics[scale=0.5, height=5.2cm]{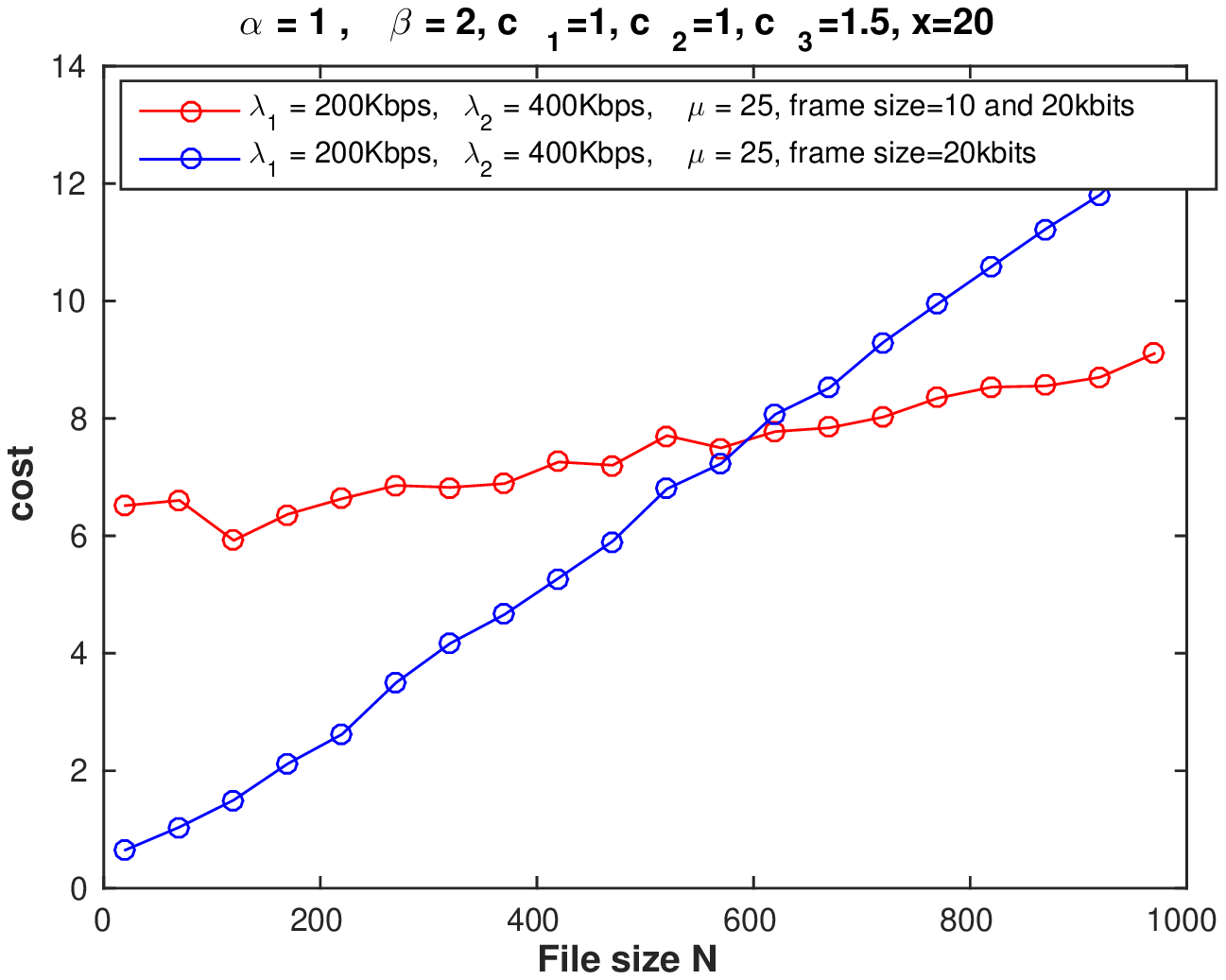}
\caption{\footnotesize{The cost for progressive streaming versus the adaptive streaming} \label{progadap2}}
\end{center}
\end{figure}

%The model can be used to optimize the QoE. Given the QoE requirements based on the number of starvation events that can annoy the users, the starvation probability which depends on the file size $ Z $ is limited by a threshold that we call $ \epsilon $. One can compute the optimum start-up threshold that satisfy the criteria $ P_s(Z) < \epsilon $ by $$ x^* = min \{ x | P_s(Z) < \epsilon \}. $$ Fig. \ref{tradeoffxps} gives the optimum start-up threshold for different file sizes, for different QoE requirements. It first shows that the optimum value of $ x $ increases with the file size, but we can also see that this value increases for systems which require small number of starvation. This is the reason why authors in \cite{buffer_starv} defined a cost to model the trade-off between the probability of starvation and the start-up delay which lead to the same results.  This figure also shows that the change on the network throughput can be benefit for the user. 
In this section, we introduce an optimization problem of the QoE by including different metrics and incorporating user preferences by associated a weight to each metric. We denote by $C(x,Z)$ the cost of a user watching the media stream,
$$ 
C(x,Z)=c_1.P_{ns}(Z)+c_2.{T_x}+c_3 .\Delta F .T
$$ 
where $P_{ns}(Z)$ is the number of starvation, $T_x$ is the start-up delay, $\Delta F$ is the lost on video quality  and $T$ is the fraction  of the total session
time spent in low bit-rate.  $c_1$, $c_2$ and $c_3$ are  depending on the user preferences of the three metrics (starvation, start-up delay and video quality).  Based on the user preferences, we compare the cost of QoE for two scenarios. In scenario 1, the  adaptive bitrate streaming is not used and for the second scenario, the adaptive bitrate streaming is used in order to adjust the quality of a video stream according the available bandwidth. 
We consider a network that throughput varies between 200Kbps and 400Kbps. For the adaptive bitrate streaming, we have two coding rates depending on the throughput. This leads to two different frame sizes (10kbits, 20kbits). Then we compute the cost $c(x,Z)$ for $x=20$ and $Z=1000$. \\
In fig. \ref{progadap} and \ref{progadap2}, we compare the cost for the two scenarios. For short video duration, the adaptive bitrate streaming is not benefit because there is a less number of starvation and the quality of the video is degraded. But, for the long video duration, the adaptive bitrate streaming becomes interesting because the low coding rate decreases the number of the starvation. In fig. \ref{progadap} and \ref{progadap2}, we can see that the value of the parameter $c_3$ changes the preference of the user for the quality of the video. For $c_3=1$, $c_3=1.5$, we use the adaptive bitrate streaming when the size of the file is more than 400, 600 frames respectively. Otherwise, the adaptive bitrate streaming is not necessary.
%\begin{figure}[ht!]
%\begin{center}
%\includegraphics[scale=0.5, height=5.2cm]{optimum4.eps}
%\caption{\footnotesize{The optimum start-up threshold for different QoE requirements} \label{tradeoffxps}}
%\end{center}
%\end{figure}


\section{Conclusion}
\label{clc}
In this paper, we have proposed a new analytical framework to compute the QoE of video streaming in the network modeled by the Markov Modulated Fluid Model. We found the probability of starvation and the start-up delay in solving Partial Differential Equations through the Laplace Transform method. This allowed us to compute the number of starvation during the video session that is an important metric of the quality of experience of the user. In addition, we have presented simulation results using ns3 to show the correctness of our model. We have proposed a method to optimize the quality of experience given a trade-off between the player starvation and the quality of the video. These results show that the adaptive bitrate streaming could impact negatively on the quality of the short video duration.

\section*{Appendix}
\label{apend}
\begin{proof}
When $ \lambda_1 \neq \mu $, $ \lambda_2 \neq \mu $, \[
\small{ Q+s(\omega)R-\omega I=\begin{bmatrix}
-\beta +(\lambda_2 -\mu )s(\omega)-\omega & \beta \\
\alpha & -\alpha +(\lambda_1 -\mu )s(\omega)-\omega
\end{bmatrix}
}\] 
$ (Q+s(\omega)R-\omega I)\phi(\omega)=0 $ and $ \phi^{k} $ is the eigenvector correspondind to $ s_k(\omega) $ according to section \ref{first_passage}, then \[
\phi^k =\begin{bmatrix}
1 & \frac{\beta+\omega -(\lambda_2 - \mu )s_k(\omega)}{\beta}
\end{bmatrix}^T,
k=0,1.
\]
When $ \lambda_1 < \mu $, $ \lambda_2 \geq \mu $, we use only $ \phi^1(\omega) $ in computing the distribution, since $ Re(s_0(\omega)) >0 $. Thus the distribution $ \tilde{H}_1(x,\omega) $ becomes \[
\begin{bmatrix}
\tilde{H}_{11}(x,\omega) \\
\tilde{H}_{21}(x,\omega)
\end{bmatrix} =a_{11}e^{s_1(\omega)x}\begin{bmatrix}
1 \\
\frac{\beta+\omega -(\lambda_2 - \mu )s_1(\omega)}{\beta}
\end{bmatrix}
\]
$ a_{11}=\tilde{H}_{11}(0,\omega) $ and $ \tilde{H}_{11}(0,\omega)=1 $ because if we start without packets in the buffer in state 1, we'll have starvation with probability 1 within the same state. So $ a_{11}=1 $, that yields the result of the theorem. The same proof holds in the case $ \lambda_2 < \mu $, $ \lambda_1 \geq \mu $ by interchanging $ \lambda_1 $ and $ \lambda_2 $. \\
When $ \lambda_1 < \mu $, $ \lambda_2 < \mu $, we use both $ \phi^0(\omega) $ and $ \phi^1(\omega) $. Thus we have 
\begin{align*}
\tilde{H}_1(x,\omega) &= \begin{bmatrix}
\tilde{H}_{11}(x,\omega) \\
\tilde{H}_{21}(x,\omega)
\end{bmatrix} \\
&= a_{11}e^{s_0(\omega)x} \begin{bmatrix}
1 \\
\frac{\beta+\omega -(\lambda_2 -\mu )s_0(\omega)}{\beta}
\end{bmatrix}+ \\
&~~~~ a_{21}e^{s_1(\omega)x} \begin{bmatrix}
1 \\
\frac{\beta+\omega -(\lambda_2 -\mu )s_1(\omega)}{\beta}
\end{bmatrix}
\end{align*}
Using the initial condition $ \tilde{H}_{11}(0,\omega) =1 $ and $ \tilde{H}_{21}(0,\omega) =0 $, we solve the following system \[
\left\{
\begin{array}{r c l}
a_{11} + a_{21} &=& 1 \\
a_{11}\frac{\beta+\omega -(\lambda_2 -\mu )s_0(\omega)}{\beta} + a_{21}\frac{\beta+\omega -(\lambda_2 -\mu )s_1(\omega)}{\beta} &=& 0
\end{array}
\right.
\]
and get
\begin{align*}
a_{11} &=\frac{\beta+\omega -(\lambda_2 - \mu )s_1(\omega)}{(\lambda_2 -\mu )(s_0(\omega)-s_1(\omega))} \\
a_{21} &=\frac{\beta+\omega -(\lambda_2 -\mu )s_0(\omega)}{(\lambda_2 - \mu )(s_1(\omega)-s_0(\omega))} 
\end{align*}
The same proof yiels for $ \tilde{H}_2(x,\omega) $.
\end{proof}

\end{document}